\documentclass [pra,twocolumn,showpacs,floatfix]{revtex4}
\usepackage{dcolumn,bm,amsmath}
\def\etal{\emph{et~al.}}
\newcommand{\eref}[1]{(\ref{#1})}
\newcommand{\Eref}[1]{Eq.~(\ref{#1})}
\newcommand{\tref}[1]{Table~\ref{#1}}
\newcommand{\rtw}{\rightarrow}
\newcommand{\cm}{cm$^{-1}$}
\begin{document}
%####################################################################
\title{Space-time variation of the fine structure constant and evolution of
isotope abundances.}
\author{M. G. Kozlov}
\email{mgk@MF1309.spb.edu}
\affiliation{Petersburg Nuclear Physics Institute, Gatchina, 188300, Russia}
\author{V. A. Korol}
\affiliation{St.Petersburg Polytechnical University, St.Petersburg, Russia}
\author{J. C. Berengut}
\author{V. A. Dzuba}
\author{V. V. Flambaum}
\affiliation{University of the New South Wales, Sydney, Australia}
\date{\today}

\begin{abstract}
At present several groups are analyzing astrophysical data in a search for
time-variation of the fine structure constant $\alpha$. Here we discuss
how to exclude systematic effects caused by the changes in the 
isotope abundances during the evolution of the universe. We suggest using
particular combinations of the transition frequencies of O~II, Al~II, Al~III, 
Si~II and Mn~II as anchors, which are insensitive to $\alpha$-variation 
and to changes in isotope abundances. These anchors can be used to 
determine the cosmological redshift. Then, one can use other 
combinations of frequencies as probes for the time-variation of 
$\alpha$ and another set as probes for the isotopic 
abundances. In this way it is possible to eliminate one 
source of the systematic errors in the search for $\alpha$-variation and 
get new information about evolution of the isotopes. On the level of accuracy 
that has already been reached in the search for $\alpha$-variation it is 
possible to see $\sim$10\% changes in isotope abundances of Si and Ni. 
\end{abstract}


\maketitle

%###############################################
\subsection*{Introduction}
%###############################################
\label{intro}

If, during previous stages of the evolution of the universe,
the fine structure constant $\alpha$ was different from its present value
$\alpha_0$, we can expect small deviations of atomic frequencies from
their laboratory values \cite{DFW99b}:
%------------------------------------------------------------
\begin{subequations}
\label{i1}
\begin{eqnarray}
  \omega  &=&  \omega_{\textrm{lab}} + q x + O(x^2)
\label{i1a}\\
  &\approx& \omega_{\textrm{lab}} + 2q\, {\Delta \alpha}/{\alpha},
\label{i1c}\\
  x &\equiv& \left(\alpha/\alpha_0\right)^2-1,
\nonumber
\end{eqnarray}
\end{subequations}
%------------------------------------------------------------
where the parameter $q$ rapidly grows with the nuclear charge $Z$. At least
three groups are looking for such deviations in the spectra of distant 
quasars \cite{WFC99,WMF01,MFW03,QRL03,SCP04}. These groups use different
data and their results are not fully consistent with each other. In
\cite{WFC99,WMF01,MFW03} a nonzero $\alpha$-variation is reported,
while \cite{QRL03,SCP04} agree with the constant $\alpha$:
%------------------------------------------------------------
\begin{eqnarray}
\label{i2}
\frac{\Delta \alpha}{\alpha}
&=& 10^{-5} \times \left\{
\begin{array}{lcr}
-0.57 \pm 0.11                    & \quad & \textrm{\cite{MFW03}}\\
-0.04 \pm 0.19 \pm 0.27_{\rm sys} & \quad & \textrm{\cite{QRL03}}\\
-0.06 \pm 0.06                    & \quad & \textrm{\cite{SCP04}}\\
\end{array}\right.
\end{eqnarray}
%------------------------------------------------------------
The frequency shifts which correspond to \Eref{i2} are of the
order of magnitude of the typical isotope shifts. Therefore, any changes
in isotope abundances can cause  systematic errors in
the search for $\alpha$-variation (see, for example, discussions in
\cite{Lev94,VPI01,MFW03,AMO04} and references therein).

Here we use recent calculations of the isotope shifts 
\cite{BDF03,BDFK03a,BDFK04} to reanalyse the role 
of the isotope evolution in the search for $\alpha$-variation.
We suggest how to exclude the possible systematic error caused by isotope 
shift from the search for $\alpha$-variation. We also show that the
spectroscopic data from \Eref{i2} can be used to get information about
evolution of isotope abundances of Si and Ni.

The shift in frequency of any transition in an isotope with mass number 
$A'$ with respect to an isotope with mass number $A$ can be expressed as:
%------------------------------------------------------------
\begin{equation}
\label{i3}
\Delta \omega^{A', A} = k_{\rm MS}
\left(\frac{1}{A'} - \frac{1}{A} \right) + F \delta \langle r^2 \rangle
    ^{A', A} \ ,
\end{equation}
%------------------------------------------------------------
where $\langle r^2 \rangle$ is the mean square nuclear radius.
The mass shift constant $k_{\rm MS}$ accounts for the normal mass shift
and for the specific mass shift \cite{Sob79}: 
$k_{\rm MS}=k_{\rm NMS} + k_{\rm SMS}$, where
%------------------------------------------------------------
\begin{equation}
k_{\rm NMS} = -\frac{\omega}{1823},
\label{i4}
\end{equation}
%------------------------------------------------------------
and the value 1823 refers to the ratio of the atomic mass unit (amu) to the
electron mass. The values of the constants $k_{\rm SMS}$ for the specific
mass shift and $F$ for the volume shift in \eref{i3} depend on the
details of the atomic structure (see \cite{BDF03,BDFK04} for details). 
Note that we use the sign convention \cite{BDF03,BDFK04} with 
$k_{\rm NMS}<0$, which differs from the convention used in \cite{Sob79} 
and some other publications.

Present accuracy of the theory for $k_{\rm SMS}$ for atoms with more 
than one valence electron is not always sufficient for our purpose. However, 
it is possible to improve the theory by applying techniques suggested in
\cite{DFK96a,DFK96b}. Note that calculations for the ions with one electron
above closed shells are already sufficiently accurate \cite{BDF03,SJ01,Tup03}. 
Below we ignore the volume shift of \Eref{i3}. For light elements it is much 
smaller than the mass shift, but becomes more important for such elements 
as Ni, Zn, and Ge. Still, its size is comparable to the accuracy to which 
we currently know the isotope shift parameters $k_{\rm MS}$.

%####################################################################
\begin{table*}[p]
\caption
{Results of the calculations of the $\alpha$-variation parameters $q$ 
\cite{DFW99a,DFK02,BDFM04,ADF04} and the isotope shift parameters $k_{\rm MS}$
\cite{BDF03,BDFK03a} for the transitions used in astrophysical surveys.
The error bars are given in parentheses where available. Column 
$\Delta \omega_\alpha$ gives the shifts that correspond to the 
$\alpha$-variation observed in \cite{MFW03}.
Column $\Delta\omega_1$ shows the line shift caused by a 10\% abundance 
transfer from the leading isotope to the next to leading (by abundance). 
Similarly, column $\Delta\omega_2$ presents the line shift for the case when 
natural abundance is substituted by a single leading isotope.}

\label{tab1}

\begin{tabular}{llclrrlrlrrrr}
\hline
\hline
 \multicolumn{1}{c}{Ion}
&\multicolumn{3}{c}{Transition}
&\multicolumn{1}{c}{$\omega_0$}
&\multicolumn{2}{c}{$q$}
&\multicolumn{2}{c}{$k_{\rm MS}$}
&\multicolumn{1}{r}{$\Delta\omega_\alpha$}
&\multicolumn{1}{r}{$\Delta\omega_1$}
&\multicolumn{1}{r}{$\Delta\omega_2$}
&\multicolumn{1}{r}{References}\\
&&&
&\multicolumn{1}{c}{(\cm)}
&\multicolumn{2}{c}{(\cm)}
&\multicolumn{2}{c}{(\cm $\cdot$ amu)}
&\multicolumn{3}{c}{($10^{-3}$\cm)}
&\\
\hline\\
O II  &$^4\!S_{3/2}^o[2s^22p^4]$&$\rtw$&$^4\!P_{5/2}[2s2p^4] $
& 119873&$  346$&      &   ---  &        
&$  -4$&  ---  &  0  & \cite{BDFM04}\\
      &          &$\rtw$&$^4\!P_{3/2}[2s2p^4] $
& 120000&$  489$&      &   ---  &        
&$  -6$&  ---  &  0  & \cite{BDFM04}\\
      &          &$\rtw$&$^4\!P_{1/2}[2s2p^4] $
& 120083&$  574$&      &   ---  &
&$  -7$&  ---  &  0  &\cite{BDFM04}\\
\\
Mg I  &$^1S_0[3s^2]$&$\rtw$&$^1P_{1}^o[3s3p]$
\footnote{For this transition we give experimental value for $k_{\rm MS}$ 
from \cite{Hal79}.}
&  35051&$   86$&      &$ -14.5$&(0.3)   
&$  -1$&     5    &$  -8 $& \cite{Hal79, DFW99a}\\
Mg II &$3s_{1/2}$&$\rtw$&$3p_{1/2}           $
&  35669&$  120$&      &$ -32.1$&(0.4)   
&$  -1$&    10    &$ -16 $& \cite{BDF03, DFW99a}\\
      &          &$\rtw$&$3p_{3/2}           $
&  35761&$  211$&      &$ -32.1$&(0.2)   
&$  -2$&    10    &$ -16 $& \cite{BDF03, DFW99a}\\
\\
Al II &$^1\!S_{0}[3s^2] $&$\rtw$&$^3\!P_{0}[3s3p] $
&  37393&$  146$& (15) &   ---  &       
&$  -2$&    ---   &   0   & \cite{ADF04}\\
      &               &$\rtw$&$^3\!P_{1}[3s3p]    $
&  37454&$  211$& (20) &   ---  &
&$  -2$&    ---   &   0   & \cite{ADF04}\\
Al III&$3s_{1/2}     $&$\rtw$&$3p_{1/2}    $
&  53682&$  216$&      &   ---  &
&$  -2$&    ---   &   0   & \cite{DFW99a}\\
      &               &$\rtw$&$3p_{3/2}    $
&  53920&$  464$&      &   ---  &
&$  -5$&    ---   &   0   & \cite{DFW99a}\\
\\
Si II &$^2\!P^o_{1/2}[3s^23p]$&$\rtw$&$^2\!D_{3/2}[3s3p^2] $
&  55309&$  520$& (30) &$  -63 $&(15)   
&$  -6$&     8    &$ -9  $& \cite{BDF03, DFK02}\\
      &                       &$\rtw$&$^2\!S_{1/2}[2s^24s] $
&  65495&$   50$& (30) &$   13 $&(20)    
&$  -1$&$   -2   $&$  2  $& \cite{BDF03, DFK02}\\
Si IV &$3s_{1/2}$             &$\rtw$&$3p_{1/2} $
&  71290&$  346$&      &$ -90.3$&(0.4) 
&$  -4$&    11    &$ -13 $& \cite{BDF03, DFW99a}\\
      &                       &$\rtw$&$3p_{3/2} $
&  71750&$  862$&      &$ -89.6$&(0.2) 
&$ -10$&    11    &$ -13 $& \cite{BDF03, DFW99a}\\
\\
Ti II &$^4\!F_{3/2}[3d^24s]$  &$\rtw$&$^4\!F^o_{3/2}[3d^24p]$
&  30837&$  541$& (50) &$ -20  $& (4) 
&$  -6$&$   -2   $&$  1  $& \cite{BDFM04,BDFK03a}\\
      &                       &$\rtw$&$^4\!F^o_{5/2}[3d^24p]$
&  30959&$  673$& (70) &$ -20  $& (4) 
&$  -8$&$   -2   $&$  1  $& \cite{BDFM04,BDFK03a}\\
\\
Cr II &$^6\!S_{5/2}[3d^5]$&  $\rtw$&$^6\!P^o_{3/2}[3d^44p]$
&  48399&$-1360$&(150) &$  63  $& (40) 
&$  16$&$   -2   $&   1   & \cite{BDFK03a, DFK02}\\
      &                     &$\rtw$&$^6\!P^o_{5/2}[3d^44p]$
&  48491&$-1280$&(150) &$  63  $& (40) 
&$  15$&$   -2   $&   1   & \cite{BDFK03a, DFK02} \\
      &                     &$\rtw$&$^6\!P^o_{7/2}[3d^44p]$
&  48632&$-1110$&(150) &$  63  $& (40) 
&$  13$&$   -2   $&   1   & \cite{BDFK03a, DFK02} \\
\\
Mn II &$^7\!S_3[3d^54s]$    &$\rtw$&$^7\!P^o_2[3d^54p]   $
&  38366&$  869$& (90) &   ---   &       
&$ -10$&  ---  &  0  & \cite{BDFM04}\\
      &        	            &$\rtw$&$^7\!P^o_3[3d^54p]   $
&  38543&$ 1030$&(100) &   ---   &
&$ -12$&  ---  &  0  & \cite{BDFM04}\\
      &        	            &$\rtw$&$^7\!P^o_4[3d^54p]   $
&  38807&$ 1276$&(100) &   ---   &
&$ -15$&  ---  &  0  & \cite{BDFM04}\\
      &                     &$\rtw$&$^7\!P^o_2[3d^44s4p] $
&  83255&$-3033$&(300) &   ---   &
&$  35$&  ---  &  0  & \cite{BDFM04}\\
      &                     &$\rtw$&$^7\!P^o_3[3d^44s4p] $
&  83376&$-2825$&(300) &   ---   &
&$  32$&  ---  &  0  & \cite{BDFM04}\\
      &                     &$\rtw$&$^7\!P^o_4[3d^44s4p] $
&  83529&$-2556$&(300) &   ---   &
&$  29$&  ---  &  0  & \cite{BDFM04}\\
\\
Fe II &$^6\!D_{9/2}[3d^64s]$&$\rtw$&$^6\!D^o_{9/2}[3d^64p]$
\footnote{The lines of Fe~II used by Quast \etal\ \cite{QRL03}.}
&  38459&$ 1330$&(150) &$  -60  $&(20) 
&$ -15$&$  -4 $&$ 2 $& \cite{BDFK03a, DFK02}\\
      &                     &$\rtw$&$^6\!D^o_{7/2} [3d^64p]$
\footnotemark[2]
&  38660&$ 1490$&(150) &$  -60  $&(20) 
&$ -17$&$  -4 $&$ 2 $& \cite{BDFK03a, DFK02}\\
      &                     &$\rtw$&$^6\!F^o_{11/2}[3d^64p]$
\footnotemark[2]
&  41968&$ 1460$&(150) &$  -63  $&(20) 
&$ -17$&$  -4 $&  2  & \cite{BDFK03a, DFK02}\\
      &                     &$\rtw$&$^6\!F^o_{9/2} [3d^64p]$
\footnotemark[2]
&  42115&$ 1590$&(150) &$  -63  $&(20) 
&$ -18$&$  -4 $&  2  & \cite{BDFK03a, DFK02}\\
      &                     &$\rtw$&$^6\!P^o_{7/2} [3d^64p]$
\footnotemark[2]
&  42658&$ 1210$&(150) &$  -60  $&(20) 
&$ -14$&$  -4 $&$ 2 $& \cite{BDFK03a, DFK02}\\
      &                     &$\rtw$&$^4\!F^o_{7/2} [3d^64p]$
&  62066&$ 1100$&(300) &$  -67  $&(40) 
&$ -13$&$  -4 $&$ 2 $& \cite{BDFK03a, DFK02}\\
      &                     &$\rtw$&$^6\!P^o_{7/2} [3d^54s4p]$
\footnotemark[2]
&  62172&$-1300$&(300) &$   67  $&(40) 
&$  15$&$   4 $&$-2 $& \cite{BDFK03a, DFK02}\\
\\
Ni II &$^2\!D_{5/2}[3d^9]$  &$\rtw$&$^2\!F^o_{7/2}[3d^84p]$
&  57080&$ -700$&(250) &$   77  $&(50) 
&$   8$&$  -4 $&  18 & \cite{BDFK03a, DFK02}\\
      &                     &$\rtw$&$^2\!D^o_{5/2}[3d^84p]$
&  57420&$-1400$&(250) &$   77  $&(50) 
&$  16$&$  -4 $&  18 & \cite{BDFK03a, DFK02}\\
      &                     &$\rtw$&$^2\!F^o_{5/2}[3d^84p]$
&  58493&$  -20$&(250) &$   77  $&(50) 
&$   0$&$  -4 $&  18 & \cite{BDFK03a, DFK02}\\
\\
Zn II &$   4s_{1/2}$        &$\rtw$&$   4p_{1/2}$  
&  48481&$ 1584$& (25) &$  -70.3$&(2.3)
&$ -18$& 3  &$ -24 $& \cite{BDF03, DFK02}\\
      &                     &$\rtw$&$   4p_{3/2}$  
&  49355&$ 2490$& (25) &$  -69.3$&(2.3)
&$ -28$& 3  &$ -24 $& \cite{BDF03, DFK02}\\
\\
Ge II &$4p_{1/2}   $        &$\rtw$&$5s_{1/2}     $
&  62403&$  607$&      &$    0.7$&(2.3)
&$  -7$& 0  &$   0 $& \cite{BDF03, DFW99a}\\
\\
\hline
\hline
\end{tabular}
\end{table*}
%####################################################################

%###############################################
\subsection*{Isotopic effects}
%###############################################

In this section we discuss the implications of recent calculations of isotope 
shifts \cite{BDF03,BDFK03a,BDFK04} for the astrophysical search for 
$\alpha$-variation. Table~\ref{tab1} presents calculated values of the
parameters $q$ and $k_{\rm MS}$ for the transitions used in astrophysical 
surveys \cite{WFC99,WMF01,MFW03,QRL03,SCP04}.
In order to estimate possible isotope shifts in astrophysics we calculate 
isotope shifts $\Delta\omega_{1,2}$ for two simple assumptions:
\begin{itemize} 
\item $\Delta\omega_{1}$ corresponds to the
case when the abundance of the leading isotope $A$ is reduced by 10\%, 
while the abundance of the next to leading isotope $A'$ is increased 
by the same amount. That means that for Si and Cr $A'=A+1$; for
Mg, Ni, and Zn $A'=A+2$; for Ti, Fe, and Ge $A'=A-2$. Then 
$\Delta\omega_{1} = 0.1 \times \Delta\omega^{A', A}$.
%is the isotope shift $\Delta\omega^{A', A}$ times $0.1$.
\item $\Delta\omega_{2}$ corresponds to the substitution of the
terrestrial abundances with the single leading isotope. For O, Al, and Mn
that means zero shifts as these elements have no isotopes.
\end{itemize}

The shifts $\Delta\omega_{1}$ and $\Delta\omega_{2}$ are usually of the 
opposite sign, and we use them to estimate the range of the possible
isotopic shifts for distant absorber. In order to compare isotopic effects
with the expected effects from $\alpha$-variation, \tref{tab1} also 
presents the shifts $\Delta\omega_\alpha$ which correspond to 
$\Delta\alpha/\alpha=-0.57\times 10^{-5}$ \cite{MFW03}. Note that our 
first assumption of a 10\% change in abundances is a rather conservative 
one. Sometimes, much more dramatic scenarios of the isotope evolution are 
considered (see, for example \cite{AMO04} and the references therein). 
For these scenarios our conclusions will only be enhanced.

\tref{tab1} shows that the shifts $\Delta\omega_{1,2}$ are indeed of the
same order of magnitude as the shifts observed in \cite{MFW03}. Moreover,
in some cases there is a strong correlation between $\Delta\omega_\alpha$ 
and $\Delta\omega_{1,2}$. That confirms earlier suggestions that isotopic 
effects may lead to significant systematic errors in the search for 
$\alpha$-variation.

Let us, for example, consider Fe~II. Here isotope shifts are relatively 
small. On the other hand, coefficients $q$ and $k_{\rm MS}$ are strongly 
correlated, and it is impossible to disentangle these two effects.
For the lines used in \cite{QRL03}, the isotope shifts 
which correspond to our two limiting cases would
imitate the following $\alpha$-variation:
%------------------------------------------------------------
\begin{eqnarray}
\label{e1}
\frac{\Delta \alpha}{\alpha}
&=& 10^{-5} \times \left\{
\begin{array}{lcr}
-0.14  & \quad & \textrm{case 1,}\\
+0.07  & \quad & \textrm{case 2.}\\
\end{array}\right.
\end{eqnarray}
%------------------------------------------------------------
These values lie within the error bars given by Quast \etal, but will not 
allow one to improve their results significantly, unless isotope effects are 
accounted for. That, in turn, will require one to include other lines in 
the analysis.

%###############################################
\subsection*{Anchors and probes}
%###############################################

When spectra of distant astrophysical objects are observed, one needs to
exclude the cosmological redshift. That can be done by looking at the
frequency ratios for different transitions, i.e.~by comparing
$(\omega_i/\omega_k)_{\rm astro}$ with $(\omega_i/\omega_k)_{\rm lab}$.
When forming these ratios it is convenient to use ``anchor'' lines, which
are not sensitive to the variation of the parameter of interest.

As we saw above, there can be two effects of the same order of magnitude, 
i.e.~possible $\alpha$-variation and variation of isotope abundances. 
Therefore, we need anchors which are insensitive to both effects. 
In \tref{tab1} there is only one line of Si~II, which may be considered
suitable as anchor:
%------------------------------------------------------------
\begin{eqnarray}
A_I &=& \omega\! \left(^2\!P^o_{1/2}\rtw {}^2\!S_{1/2}\right)_{\rm Si\,II}
= 65495\textrm{~\cm}.
\label{d2}
\end{eqnarray}
%------------------------------------------------------------
It is preferable to have more than one anchor. However, all other
lines have either relatively large $q$, or $k_{\rm MS}$, or both.
Aluminum, oxygen and manganese ions have only one stable isotope 
each (leading isotope abundance for $^{16}\!$O is greater than $99\%$). 
In order to exclude the $\alpha$-dependence of transition frequencies we 
can use the following combinations of $^{16}$O and $^{27}\!$Al:
%------------------------------------------------------------
\begin{eqnarray}
\label{vk3}
A_{II}
&=&0.62\cdot\omega\!\left(^4\!S_{3/2}^o \rtw{}^4\!P_{5/2}\right)_{\rm O\,II}
\\
&-&0.38\cdot\omega\!\left(^4\!S_{3/2}^o \rtw{}^4\!P_{1/2}\right)_{\rm O\,II}
= 29650\textrm{~\cm},
\nonumber \\
\label{vk1}
A_{III}
&=& 0.59 \cdot \omega\!\left(3s_{1/2} \rtw 3p_{1/2}\right)_{\rm Al\,II}
\\
&-& 0.41 \cdot \omega\!\left(3s_{1/2} \rtw 3p_{3/2}\right)_{\rm Al\,II}
= 6781\textrm{~\cm},
\nonumber \\
\label{d3}
A_{IV}
&=& 0.68 \cdot \omega\!\left(3s_{1/2} \rtw 3p_{1/2}\right)_{\rm Al\,III}
\\
&-& 0.32 \cdot \omega\!\left(3s_{1/2} \rtw 3p_{3/2}\right)_{\rm Al\,III}
= 19465\textrm{~\cm}.
\nonumber
\end{eqnarray}
%------------------------------------------------------------
For $^{55}$Mn we can form several anchors, such as:
%------------------------------------------------------------
\begin{eqnarray}
\label{vk2}
A_{V}\!
&=\!& 0.67 \cdot \omega\! \left(^7\!S_{3} \rtw{}^7\!P^o_4[3d^5 4p]
\right)_{\rm Mn\, II}
\\
&+\!& 0.33 \cdot \omega\! \left(^7\!S_{3} \rtw{}^7\!P^o_4[3d^4 4s 4p]
\right)_{\rm Mn\, II}\!
= 53699\textrm{~\cm}\!.
\nonumber
\end{eqnarray}
%------------------------------------------------------------
Numerical factors in \eref{vk3} ---~\eref{vk2} are formed from the 
$q$-factors for two transitions and are normalized to unity. In the 
leading order $A_{II}$~---~$A_{V}$ do not depend on $\alpha$ and can 
be used as anchors.

Other elements in \tref{tab1} have more than one isotope.
Strong correlation between the factors $q$ and
$k_{\rm MS}$ for Fe~II allows us to form several combinations of frequencies
suitable as anchors. It could be possible to make anchors from practically
any pair of lines in Fe~II which include the line 62172~\cm, for example:
%------------------------------------------------------------
\begin{eqnarray}
\label{d4}
A_{VI}
&=& 0.5\cdot\omega\!\left(^6\!D_{9/2} \rtw {}^6\!D^o_{9/2}\right)_{\rm Fe\,II}
\nonumber \\
&+& 0.5\cdot\omega\!\left(^6\!D_{9/2} \rtw {}^6\!P^o_{7/2}\right)_{\rm Fe\,II}
= 50316\textrm{~\cm}.
\nonumber
\end{eqnarray}
%------------------------------------------------------------
%%The problem here is in the low accuracy of the available calculations for 
%%Fe~II. Therefore, before more accurate calculations are made it may be 
%%better to avoid using lines of Fe~II as anchors.

Now, when we have several anchors to exclude the cosmological redshift, we
can look for the combinations of frequencies which are sensitive to only
the mass shift or to the variation of $\alpha$. These combinations can be 
used as ``probes'' of isotope abundances and $\alpha$-variation. For example,
the combination
%------------------------------------------------------------
\begin{eqnarray}
\label{d5}
P_{I}
&=& 0.64 \cdot \omega\!\left(3s_{1/2} \rtw 3p_{1/2}\right)_{\rm Mg\,II}
\\
&-& 0.36 \cdot \omega\!\left(3s_{1/2} \rtw 3p_{3/2}\right)_{\rm Mg\,II}
= 9740\textrm{~\cm}
\nonumber 
\end{eqnarray}
%------------------------------------------------------------
is insensitive to $\alpha$-variation and can serve as a probe of the
abundances of Mg isotopes in distant absorbers.

Note that even though $q$-factors for Mg are relatively small, it is still
necessary to use $P_I$ as a probe, rather than individual lines. Since
$\alpha$-variation on the scale of $\Delta\alpha/\alpha \approx -0.6 \times
10^{-5}$ is currently not excluded, we can expect the shift $\delta
\omega(3s_{1/2} \rtw 3p_{3/2}) \approx -0.002$~\cm, which is comparable to 
the current sensitivity of observations to the frequency shifts.

By analogy with Mg~II it is possible to form combinations for the fine
structure doublets in Si~IV and Zn~II:
%------------------------------------------------------------
\begin{eqnarray}
\label{d7}
P_{II}
&=& 0.71 \cdot \omega\!\left(3s_{1/2} \rtw 3p_{1/2}\right)_{\rm Si\,IV}
\\
&-& 0.29 \cdot \omega\!\left(3s_{1/2} \rtw 3p_{3/2}\right)_{\rm Si\,IV}
= 30381\textrm{~\cm},
\nonumber \\
\label{d7a}
P_{III}
&=& 0.61 \cdot \omega\!\left(4s_{1/2} \rtw 4p_{1/2}\right)_{\rm Zn\,II}
\\
&-& 0.39 \cdot \omega\!\left(4s_{1/2} \rtw 4p_{3/2}\right)_{\rm Zn\,II}
= 10423\textrm{~\cm}.
\nonumber 
\end{eqnarray}
%------------------------------------------------------------

For most other transitions in \tref{tab1} there is strong
correlation between parameters $q$ and $k_{\rm MS}$, and we cannot
exclude the dependence on $\alpha$. The only exception is Ni~II,
where the transition with frequency 58493~\cm\ is insensitive to
$\alpha$-variation and two other transitions have different parameters $q$.
That gives us two more probes of isotope abundances:
%------------------------------------------------------------
\begin{eqnarray}
\label{d9}
P_{IV}
&=& \omega\!\left({}^2\!D_{5/2} \rtw {}^2\!F^o_{5/2}\right)_{\rm Ni\,II}\!
= 58493\textrm{~\cm},
\\
\label{d9a}
P_{V}
&=& 0.67 \cdot
\omega\!\left({}^2\!D_{5/2} \rtw {}^2\!F^o_{7/2}\right)_{\rm Ni\,II}
\\
&-& 0.33 \cdot
\omega\!\left({}^2\!D_{5/2} \rtw {}^2\!D^o_{5/2}\right)_{\rm Ni\,II}\!
= 18952\textrm{~\cm}.
\nonumber 
\end{eqnarray}
%------------------------------------------------------------

\begin{table}[tbh]
\caption
{$\alpha$-variation probes, insensitive to isotope shifts.}

\label{tab2}

\begin{tabular}{lrrr}
\hline
\hline
\multicolumn{1}{c}{Ion}
&\multicolumn{1}{c}{$\omega$}
&{$q~\ $}
&\multicolumn{1}{c}{Frequency}\\
&{\cm}
&{\cm}
&\multicolumn{1}{c}{combinations}\\
\hline

O II  & 119873 &$  346$&$ \omega(^4\!S_{3/2}^o\rtw{}^4\!P_{5/2})$\\
      & 120000 &$  489$&$ \omega(^4\!S_{3/2}^o\rtw{}^4\!P_{3/2})$\\
      & 120083 &$  574$&$ \omega(^4\!S_{3/2}^o\rtw{}^4\!P_{1/2})$\\
\hline

Al II &  37393 &$  146$&$ \omega(^1\!S_{0}\rtw{}^3\!P_{0})$\\
      &  36454 &$  211$&$ \omega(^1\!S_{0}\rtw{}^3\!P_{1})$\\
%%\hline
Al III&  53682 &$  216$&$ \omega(s_{1/2}\rtw 3p_{1/2})$\\
      &  53920 &$  464$&$ \omega(s_{1/2}\rtw 3p_{3/2})$\\
\hline

Si IV &    230 &$  258$&$ 1/2\cdot\omega(3s_{1/2}\rtw 3p_{3/2})$\\
      &        &       &$-1/2\cdot\omega(3s_{1/2}\rtw 3p_{1/2})$\\
\hline

Mn II &  38366 &$  869$&$ \omega(^7\!S_3\rtw{}^7\!P_2[3d^54p])$\\
      &  38543 &$ 1030$&$ \omega(^7\!S_3\rtw{}^7\!P_3[3d^54p])$\\
      &  38807 &$ 1276$&$ \omega(^7\!S_3\rtw{}^7\!P_4[3d^54p])$\\
      &  83255 &$-3033$&$ \omega(^7\!S_3\rtw{}^7\!P_2[3d^44s4p])$\\
      &  83376 &$-2825$&$ \omega(^7\!S_3\rtw{}^7\!P_3[3d^44s4p])$\\
      &  83529 &$-2556$&$ \omega(^7\!S_3\rtw{}^7\!P_4[3d^44s4p])$\\
\hline

Ni II &    170 &$ -350$&$ 1/2\cdot\omega(^2\!D_{5/2}\rtw{} ^6\!D^o_{5/2})$\\
      &        &$     $&$-1/2\cdot\omega(^2\!D_{5/2}\rtw{} ^6\!F^o_{7/2})$\\
      &    537 &$  690$&$ 1/2\cdot\omega(^2\!D_{5/2}\rtw{} ^6\!F^o_{5/2})$\\
      &        &$     $&$-1/2\cdot\omega(^2\!D_{5/2}\rtw{} ^6\!D^o_{5/2})$\\
\hline

Zn II &    437 &$  453$&$ 1/2\cdot\omega(4s_{1/2}\rtw{} 4p_{3/2})$\\
      &        &$     $&$-1/2\cdot\omega(4s_{1/2}\rtw{} 4p_{1/2})$\\
\hline

Ge~II &  62403 &$  607$&$ \omega(4p_{1/2}\rtw 5s_{1/2})$\\

\hline
\hline
\end{tabular}
\end{table}
%####################################################################

The same method can be applied for building combinations that are
insensitive to isotope shift and can serve as probes for
the variation of $\alpha$. Above we suggested that several
combinations of spectral lines of O, Al, and Mn ions can be used as
anchors. These lines themselves can also be introduced as
$\alpha$-dependent probes. Indeed, these elements have only one stable
isotope each and therefore have no isotope shifts. Here, manganese is the
best probe, as it has the biggest absolute value of $q$.
Ge~II is also a good probe because of a very low value of $k_{\rm MS}$ of the
only transition $4p_{1/2}\rtw 5s_{1/2}$, listed in \tref{tab1}.

Other ions from \tref{tab1} are sensitive to isotope shifts. Still
it is possible to make several combinations to exclude this dependence.
Mg~II, Si~II, and Ti~II have relatively low values
of $q$ and would not make good probes.
For remaining ions $k_{\rm MS}$ is practically the same for all 
transitions of interest. Hence, we can form a probe by taking the difference
of any two frequencies. The results are given in \tref{tab2},
where we introduced normalization factors as above.

For Fe~II and Cr~II it is impossible to exclude isotope dependence 
because of the correlations between $q$ and $k_{\rm MS}$. In both cases 
isotopic effects are relatively small. That allows one to use these ions to
test $\alpha$-variation on the level reported in \cite{MFW03}, but not on 
the level of the result \cite{SCP04}. 
%########################################################################

%###############################################
\subsection*{Discussion of sensitivity}
%###############################################

Let us use \Eref{i2} and \tref{tab1} to estimate typical sensitivity to the
frequency shifts in modern astrophysical studies. The paper \cite{QRL03} is
based only on the analysis of the lines of Fe~II. From \tref{tab1} we see
that for all transitions in this ion $|q| \approx 1300$~\cm. \Eref{i1c}
then gives statistical accuracy $\Delta \omega \approx 0.005$~\cm. Two
other groups \cite{MFW03,SCP04} use more ions, but Fe~II is present in both
samples and we can again use the same value of $|q|$ as typical. That
leads to the estimates of the accuracy which are roughly 2 \cite{MFW03} and
3 \cite{SCP04} times better than above. This is statistical
accuracy, which is achieved by averaging results for many transitions, so
we will use the following estimate of the achievable accuracy for the
frequency shifts in astrophysical observations:
%------------------------------------------------------------
\begin{equation}
\Delta \omega \approx 0.005\textrm{~\cm}.
\label{d1}
\end{equation}
%------------------------------------------------------------

Now we estimate frequency shifts for the probes $P_I$ --~$P_V$ and compare
them with current sensitivity \eref{d1}. To do this we need to specify the
way in which isotope abundances are changed. We assume that the intensity
of the line of the leading isotope is transferred to the next to leading
one. If there are two comparable weaker lines, we assume that they are
equally increased. Within this simple model we calculate the shifts of the
centers of gravity of the lines and the resulting shift of the probe.
Comparison of this shift to the modern frequency sensitivity \eref{d1}
gives us the sensitivity to the isotope abundances. 

Mg has three stable isotopes $A=$ 24, 25, and 26 with modern abundances
$79:10:11$. Suppose that in the early universe it was
$(79-x):(10+x/2):(11+x/2)$. Using factors $k_{\rm MS}$ from \tref{tab1}
and ignoring the volume shift and saturation effects, we can estimate 
corresponding shift of the center of gravity of the probe \eref{d5}:
%------------------------------------------------------------
\begin{eqnarray}
\delta P_{I}
&=& -0.27 \frac{3 k_{\rm MS}\,x}{200 A^2}
\approx 0.00023\, x \textrm{~\cm}.
\label{s1}
\end{eqnarray}
%------------------------------------------------------------

For Si and Zn there are also two comparable admixtures to the leading
isotope. If we assume $(92-x):(5+x/2):(3+x/2)$ abundances for Si and
$(49-y):0:(28+y/2):4:(19+y/2):0:1$ abundances for Zn, we get:
%------------------------------------------------------------
\begin{eqnarray}
\delta P_{II}
&=& -0.43 \frac{3 k_{\rm MS}\,x}{200\, 28^2}
\approx 0.00074\, x \textrm{~\cm}.
\label{s2}\\
\delta P_{III}
&=& -0.22 \frac{3 k_{\rm MS}\,y}{100\, 64^2}
\approx 0.00011\, y \textrm{~\cm}.
\label{s3}
\end{eqnarray}
%------------------------------------------------------------
Ni has only one dominant admixture to the leading isotope, so we assume
$(68-x):0:(26+x):1:4:0:1$ abundances for Ni. That gives:
%------------------------------------------------------------
\begin{eqnarray}
\delta P_{IV}
&=& -\frac{2 k_{\rm MS}\,x}{100\, 58^2}
\approx 0.00046\, x \textrm{~\cm},
\label{s4}
\end{eqnarray}
%------------------------------------------------------------
while the probe $P_V$ is three times less sensitive.

Comparing \eref{s1} --~\eref{s4} with \eref{d1} we see that current 
sensitivity allows detection of a  22\% change in the abundance of the 
isotope $^{24}$Mg, a 7\% change for the isotope $^{28}$Si,
45\% for $^{64}$Zn, and an 11\% change in the abundance of $^{58}$Ni.
%------------------------------------------------------------

%############################################################
\subsection*{Conclusions}
%############################################################

In this paper we analyze the sensitivity of modern astrophysical surveys 
that search for $\alpha$-variation to isotopic effects and come to the 
following conclusions:
\begin{itemize}
\item Isotopic frequency shifts are generally likely to be of the same
order of magnitude as the present experimental sensitivity. Therefore,
isotope shifts are an important source of systematic errors.
\item It is possible to eliminate isotopic effects by taking proper 
combinations of transition frequencies. These combinations depend on the
calculated mass shift coefficients. At present not all of them are known 
with a sufficient accuracy. Corresponding theory can be improved by 
applying the method suggested in \cite{DFK96a,DFK96b}. New experimental
data on the isotope shifts is also highly desirable.
\item As a by-product of the search for $\alpha$-variation it is possible
to get information about isotopic evolution of several elements.
\item The accuracy of the surveys, which use only lines of Fe~II is
limited by the isotopic effects on the level 
$\delta \alpha/\alpha \sim 10^{-6}$. This estimate is based on a rather
conservative assumptions about possible isotope evolution.
\end{itemize}

%\medskip

We thank S.~A.~Levshakov for very helpful comments on the manuscript.
M.K. is grateful to M.~Murphy, and D.~A.~Varshalovich for valuable 
discussions. This work is supported by Russian Foundation for Basic Research,
grant No.~02-0216387, and the Australian Research Council.

%#####################################################
%\bibliographystyle{apsrev}
%\bibliography{julia_w,my_ref_w,alpha}
%#####################################################

\end{document}